\documentclass[aps,twocolumn,amssymb,floats,prb,amsmath,superscriptaddress,showpacs,a4paper]{revtex4}
\usepackage{graphicx}

\begin{document}

\title{Intersite $4p$-$3d$ hybridization in cobalt oxides: a resonant x-ray emission spectroscopy study
}

\author{Gy\"orgy Vank\'o}
\affiliation{KFKI Research Institute for Particle and Nuclear Physics, P.O. Box 49, H-1525 Budapest, Hungary}
\affiliation{European Synchrotron Radiation Facility, B.P. 220 F-38043 Grenoble Cedex 9, France }
\author{Frank M.\ F.\ de Groot}
\affiliation{Inorganic Chemistry and Catalysis, Department of Chemistry, Utrecht University, Sorbonnelaan 16, 3584 CA Utrecht, Netherlands}
\author{Simo Huotari}
\affiliation{European Synchrotron Radiation Facility, B.P. 220 F-38043 Grenoble Cedex 9, France }
\author{R. J. Cava}
\affiliation{Department of Chemistry, Princeton University, Princeton, NJ 08544, USA}
\author{Thomas Lorenz}
\affiliation{{II.} Physikalisches Institut, Universit\"at zu K\"oln, Z\"ulpicher Strasse 77, 50937 K\"oln, Germany}
\author{M. Reuther}
\affiliation{{II.} Physikalisches Institut, Universit\"at zu K\"oln, Z\"ulpicher Strasse 77, 50937 K\"oln, Germany}

\begin{abstract}

We report on strong dipole transitions to $3d$ orbitals of neighboring Co atoms in the Co $1s$ x-ray absorption pre-edge. They are revealed by applying high-resolution resonant x-ray emission spectroscopy (RXES) to compounds containing CoO$_6$ clusters. When contrasted to quadrupole local $1s3d$ excitations, these non-local $1s3d$ transitions are identified by their energy dispersion and angular dependence, their sensitivity to second-shell effects (i.e. the connection mode of the CoO$_6$ octahedra and the bond lengths), and an upwards energy shift of 2.5 eV due to the poorer screening of the core hole. The experiment reveals that the intensity of the non-local transitions gauges the oxygen-mediated $4p$-O-$3d$ intersite hybridization.  
 We propose a revised interpretation of the pre-edges of transition metal compounds. Detailed analysis of these new features in the pre-edge offers a unique insight in the oxygen mediated metal-metal interactions in transition metal-based systems, which is a crucial aspect in orbital ordering and related electronic and magnetic properties.
In addition, the exceptional resolving power of the present $1s2p$ RXES experiment allows us to demonstrate the coherent second-order nature of the underlying scattering process.

\end{abstract}

\date{\today}


\maketitle

The nature of correlation effects and the peculiarities of the many body electronic structure of transition metal oxides remain at the forefront of current condensed matter research. Some of the fundamental issues are well illustrated by the case of LaCoO$_3$, which has been studied for decades due to the fact that temperature and doping effects lead to remarkable transport and magnetic properties; however, yet even the spin state of Co is still debated for this prototype cobaltate.\cite{haverkort2006, phelan2006, vanko2006prb}
It is well-known that oxygen mediated $d$-$d$ interactions between neighboring metal sites have a crucial impact on transport, magnetic and electronic properties of oxides with strong electron correlation.\cite{goodenoughBook, khomskii1997} In contrast to this, however, the interactions between the $p$ and $d$ orbitals of neighboring metal atoms remain poorly understood. This $p$-$d$ hybridization is expected by symmetry and band-structure arguments, and is predicted by density of states (DOS) calculations.\cite{elfimov1999, joly1999, shukla2006} 
In the present work we show that novel synchrotron-based x-ray spectroscopy can reveal and characterize this intersite hybridization.

Formed from states near the Fermi-level, the oxygen-mediated $4p$-$3d$ intersite hybridization creates low-lying empty states. Band structure calculations predict that they overlap in energy with the $3d$ empty states.\cite{elfimov1999, joly1999, shukla2006} K-edge x-ray absorption spectroscopy (XAS) is capable of probing transitions to empty states of $p$ and $d$ symmetry, and thus we can expect the observation of features related to this hybridization among the lowest energy excitations. Their identification requires, however, the complete understanding of the XAS spectrum in the relevant region, i.e., at and below the main edge.

The $3d$ metal K edge typically consists of a small pre-edge due to the weak $1s3d$ quadrupole transitions plus the main $1s4p$ edge due to transitions to the empty bands of $p$-character. In fact, what XAS offers is the possibility to probe the response of the many body electronic structure upon the creation of a $1s$ core hole. The interpretation of the pre-edge is based on the quadrupole transition from the $3d^N$ ground state to the $1s^13d^{N+1}$ final state.\cite{FdGKotaniBook} 
The charge-transfer multiplet (CTM) theory was shown to give a detailed description of this region: not only does it for the quadrupolar features, but it is also able to describe local $3d$-$4p$ mixing (if any), and thus the dipolar contribution to the pre-edges for geometries that deviate substantially from inversion symmetry \cite{arrio2000,westre1997, degroot2001}.
Anomalous dipole excitations in this XAS region have been observed in a few cases,\cite{poumellec1991pss} and they were assigned to transition to the $3d$ states of neighboring metal sites.\cite{cabaret1999,shirley2004,uozumi1992} Such \emph{non-local $1s3d$ transitions} are associated with the $4p$-O-$3d$ intersite hybridization, and are the focus of our study.

A major limitation of K edge x-ray absorption studies is that the fine details of the XAS spectral response are broadened by the $1s$ core-hole lifetime broadening of more than 1 eV. Moreover, in normal XAS, the tail of the edge overlays the small pre-edge features, complicating their analysis.
An emerging technique based on the combination of absorption and emission spectroscopy, resonant x-ray emission spectroscopy (RXES, or resonant inelastic x-ray scattering, RIXS) can overcome the problems of broadening and overlap, and can unveil the details of the underlying transitions.\cite{kotani2001,krisch1995,glatzel2005ccr,shukla2006,vanko2006jpcb}

In this paper we present a detailed study of cobalt oxides utilizing $1s2p$ RXES. We give experimental evidence for the existence of the theoretically expected low-energy non-local $1s3d$ transition,\cite{elfimov1999, joly1999, shukla2006} which is a manifestation of the interplay of structural and electronic degrees of freedom in cobaltates. On the basis of these results we propose a general interpretation for the pre-edge region in XAS of transition metal compounds that differs from the accepted view of quadrupole pre-edge plus dipole edge.

\begin{figure}  
\includegraphics[width=8.2cm]{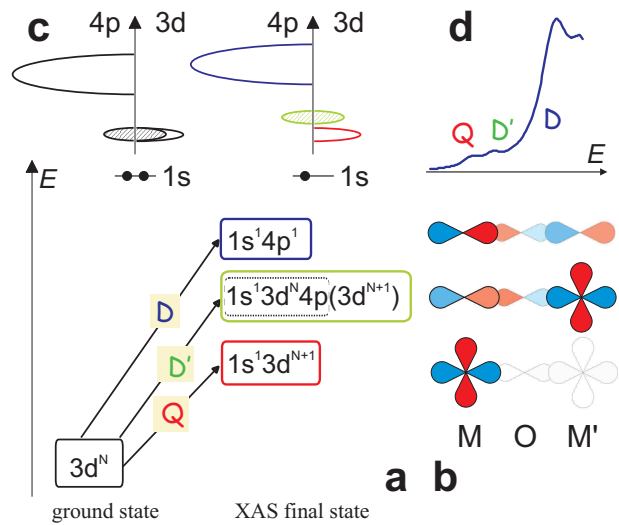}
\caption{\label{fig1} (Color online). 
(a) Schematic representation of the transitions at the K edge; Q stands for quadrupole, D for dipole, and D' for the non-local $1s3d$ dipole transitions. 
(b) The participating empty $3d$ or $4p$ metal atomic orbitals and O $2p$ orbitals are plotted for each transition. The colors illustrate the role of the orbitals, the electron is excited to darkest ones. 
(c) The ground state empty DOS and the final state densities are displayed, with the $1s$ occupation indicated. The lowest-lying empty states can be of $d$ or intersite $p$-$d$ types, which overlap in the ground state. In the XAS final states these are shifted apart due to the different screening of the $1s$ core hole.
(d) The described features are shown in an oversimplified model XAS.}
\end{figure}

The revised interpretation is illustrated in Fig.\ \ref{fig1}, from both band structure and bonding perspectives. The main edge is made up by $1s4p$ dipolar excitations (D). The lowest-lying features correspond to $1s3d$ quadrupolar transitions (Q), which can be described with the charge-transfer multiplet (CTM) approach.\cite{thole1985a,degroot1990,degroot2001,kotani2001,arrio2000,westre1997,FdGKotaniBook} In between, dipole excitations (D') appear due to excitations to second-shell metal $3d$ states, which proceed through the local $4p$ linked to the neighboring metal via the M($4p$)--O($2p$)--M'($3d$) mixing in oxides.\cite{elfimov1999,farges2005,cabaret1999} Although in the ground state the states created by oxygen-mediated $4p$-$3d$ intersite hybridization have the similar energy to the $3d$ states, the difference in core-hole screening separates the local $1s3d$ and non-local $1s3d$ contributions in the XAS spectra. The intensity of the non-local $1s3d$ excitation gives direct quantitative information on the strength of the oxygen-mediated $4p$-$3d$ intersite hybridization, which is an important property of the physics of these oxides.

In order to reveal the detailed pre-edge features, an extreme resolution is required. For this purpose we have set up a new spectrometer to perform Co $1s2p$ RXES with unprecedented resolution. The essential idea of the RXES experiment is to tune the incident energy to the $1s$ pre-edge and edge resonances and to analyze the x rays emitted in the $2p\rightarrow1s$ decay channel, with a crystal spectrometer. For a $3d^N$ ground state, the pre-edge RXES process leads via $1s^13d^{N+1}$ intermediate states to $2p^53d^{N+1}$ final states. These final states are identical to the final states in $L_{2,3}$-edge XAS, which explains the comparable information content and the improved resolution.\cite{caliebe1998,glatzel2005ccr} 
The experiment was performed at the beamline ID16 of the European Synchrotron Radiation Facility. The incident beam was monochromatized with a cryogenically cooled Si(111) double-crystal monochromator, followed by a secondary Si(440) channel-cut crystal to provide an energy bandwidth of 150--200 meV. The experimental setup was that of a Rowland-circle spectrometer with a horizontal scattering plane, operated at 90$^{\circ}$ scattering angle using a spherically bent Si(531) analyzer with a bending radius of 1m or 2m. The latter has smaller strains from the bending\cite{collart2005} allowing us to reach an overall energy resolution of 0.3 eV (FWHM); this was used for all but one (Fig. 3a) RXES spectra presented here. $1s2p_{3/2}$ RXES scans were obtained by taking the emission spectra for each incident energy (4-8 points per eV, depending on the resolution); the results are presented as contour plots. 

\begin{figure}  
\includegraphics[width=8.5cm]{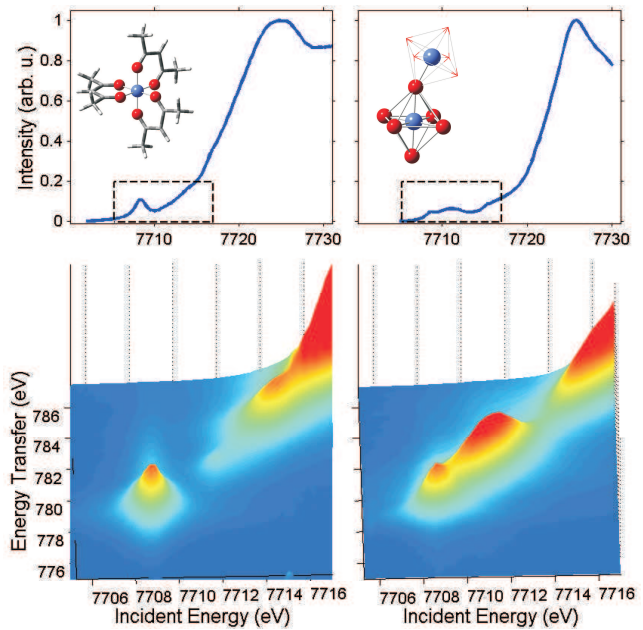}
\caption{\label{fig2} (Color online). {\em Top:} Total-fluorescence-yield detected $1s$ XAS of Co(acac)$_3$ (left) and of EuCoO$_3$ (right). {\em Bottom:} Surface plots of $1s2p$ RXES on Co(acac)$_3$ (left), and on EuCoO$_3$ (right). }
\end{figure}

\begin{figure}  
\includegraphics[width=8.5cm]{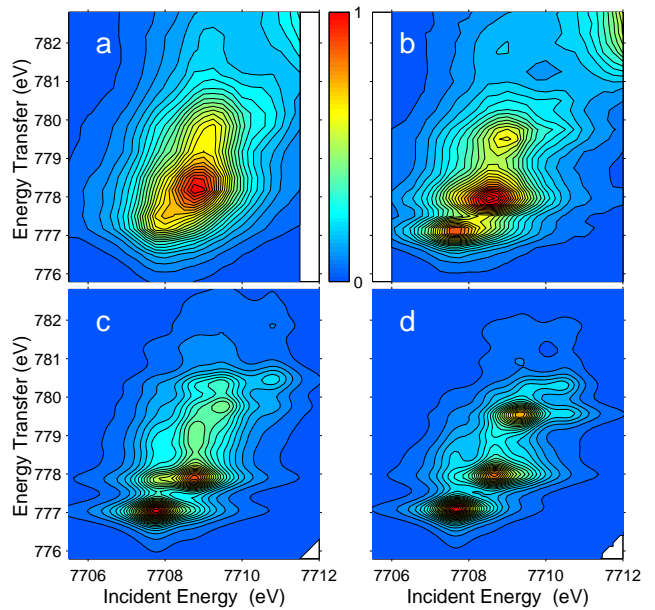}
\caption{\label{fig3} (Color online). Experimental and calculated $1s2p$ RXES spectra of CoO. Top panels show spectra taken with 1~eV~(\textbf{a}) and 0.3~eV~(\textbf{b}) overall energy resolution. Bottom panels display the modeled RXES without (\textbf{c}) and with interference (\textbf{d}) between the absorption and emission process. The relative spectral intensity is shown on the color bar between the top panels.}
\end{figure}

Before proceeding to the detailed study of the non-local $1s3d$ excitations, we show the advantages of RXES over XAS in studying the pre-edge region on examples that will be relevant to the later discussions. Firstly, the pre-edge region is better separated from the tail of the main edge in RXES than in XAS, as it is seen in Fig.\ \ref{fig2}, which presents two low-spin Co$^{3+}$ compounds. As CTM calculations and experiment show,\cite{vanko2006jpcb} the XAS of a (quasi)octahedral low-spin $3d^6$ system should consist of a single quadrupolar resonance in the preedge.\cite{Note2Eg} This is found for the case of molecular Co(acac)$_3$ (acac: acetylacetonate), as it is displayed in Figure \ref{fig2}. However, a large second structure shows up in the spectra of EuCoO$_3$, which has neighboring Co ions. 
(Below, we will argue that the reason is the relevant $4p$-$3d$ Co--O--Co intersite mixing, here we just wish to show that the pre-edge is better separated from the tail of the main edge in RXES. Exceptionally, in Fig. 2 we show the RXES spectra as colored surface plots to facilitate the understanding of the contour plot versions seen in Figs. 4a and 5.)
Secondly, the different possible features making up the pre-edge not only differ in incident energy but they are also separated along the second dimension of the RXES plane, the energy transfer. This way the large lifetime broadening of the $1s$ core hole is partly eliminated, and the spectral features are better resolved. This is especially important when the resonance energies of different sites are close, as they are in Co$_3$O$_4$,\cite{Co3O4} or when there is a rich multiplet structure in the pre-edge, such as for the case of CoO. Figure\ \ref{fig3} presents the $1s2p$ RXES of CoO (for the Co $1s$ XAS see e.g. Ref.\ \onlinecite{wu2004}). The capabilities of the technique at its new performance limits can be appreciated in this figure, which shows how the multiplet structure is resolved: the spectrum in Fig.\ \ref{fig3}a is already of good quality by current standards. However, the improvement in Fig.\  \ref{fig3}b is conspicuous: removing most of the smearing effect of the analyzer, the horizontal and vertical broadenings are determined by the corresponding ($1s$ and $2p$) core-hole lifetimes, and thus fine details of the structure are revealed.

Contrasting the obtained well-resolved experimental data with theory brings up another important aspect, which is well-known in soft x-ray spectroscopy, but has never been observed with hard X rays: the \emph{interference between the absorption and emission processes}. The calculated spectra are shown in Fig.\  \ref{fig3}, below the experimental ones. The RXES spectrum of CoO has been calculated using the CTM approach with a quadrupole transition from $3d^7$ to $1s^13d^{8}$ coupled to a dipole decay from $1s^13d^{8}$ to $2p^53d^{8}$.\cite{NoteCT} Nevertheless, charge-transfer effects have been neglected to simplify the assignments, for it can be shown that the charge-transfer parameters of CoO do not visibly modify the RIXS spectral shapes.\cite{FdGKotaniBook} The $K$ pre-edge consists of multiplet features spreading over 5 eV due to the various states of $1s^13d^{8}$ symmetry, which is completely determined by the multiplet and crystal-field effects on the $3d^{8}$ configuration.
The theoretical spectrum given on the left is determined from the multiplication of the $1s3d$ x-ray absorption intensities with the corresponding $1s2p$ x-ray emission intensities. In contrast, the calculated spectrum on the right contains the matrix elements of excitation and decay within the Kramers-Heisenberg formula.\cite{kotani2001} The difference between the two calculations is the inclusion of interference effects. 
Apart from the resolution, there is a reasonable agreement between structures in Figs.\ \ref{fig3}a and \ref{fig3}c, the less resolved experiment and the calculations without interference, which makes us recall a former assumption that interference effects are not important for $1s$ core-hole intermediate states.\cite{degroot2001} 

The high-resolution RXES spectrum, in particular the feature at around 780 eV energy transfer, is however much better simulated if interference effects are included. The excellent agreement of the structures in Figs.\ \ref{fig3}b and \ref{fig3}d proves that the effects of interference affect the spectral shape. 
This is a clear demonstration that \emph{RXES has a coherent second-order nature}.
In order for these coherence effects to be visible one should reach the same $2p$-hole final state via two or more $1s$-hole intermediate states. Mainly because the strong $2p3d$ multiplet effects in the final state, this type of interfering channels should occur frequently in the $1s2p$ RXES experiments.

Returning to the study of the $4p$-O-$3d$ intersite hybridization, we first describe our experimental strategy to tune this interaction.
In order to provide a conclusive detailed investigation of pre-edges and search for excitations with a non-local $1s3d$ character, we selected several compounds with serious restrictions on the spin and charge degrees of freedom, and structural properties: all but one subject compound have the same valence and spin state, and the local structure is always a CoO$_6$ octahedron. The main difference is the connection mode of these octahedra, which ranges from none to edge or corner sharing. The Co($4p$)--O($2p$)--Co'($3d$) mixing is most sensitive to the Co--O bond length and the Co--O--Co angle; optimal mixing is expected at short bond length and linear Co--O--Co arrangement, where the involved $2p$ orbital of the bridging oxygen can make the most effective link.
 
CoO has a rocksalt structure, with linear and orthogonal Co--O--Co geometry, the former being optimal for the $4p$-O-$3d$ intersite mixing. However, in contrast to the other compounds in this paper, it has \textit{long Co--O bonds} (2.13\AA). As we saw in Fig.\ \ref{fig3}, the rich structures of the RXES spectra are entirely understood in terms of many-body effects arising from the $1s3d$ quadrupolar transition (Q in Fig.\ \ref{fig1}); the excellent agreement of the measured and calculated spectral shapes in Fig.\ \ref{fig3}\textbf{b} and  \ref{fig3}\textbf{d} clearly shows the \emph{absence of significant non-local $1s3d$ transitions}.

\begin{figure}  
\includegraphics[width=8.2cm]{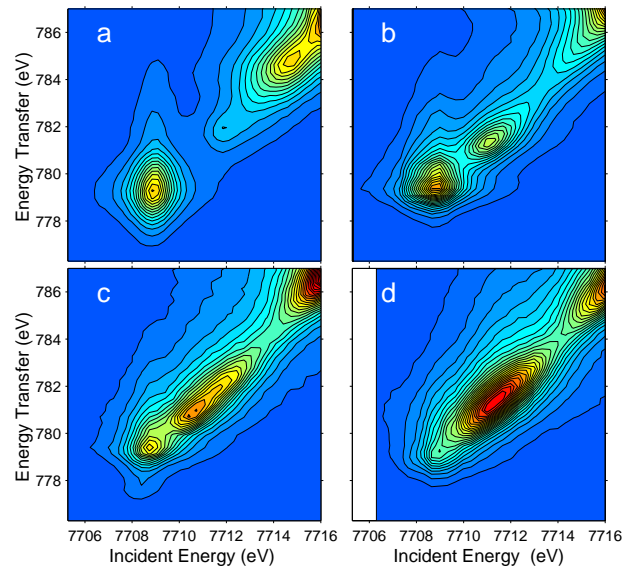}
\caption{\label{fig4} (color online). $1s2p$ RXES on (a) Co(acac)$_3$, (b) LiCoO$_2$, (c) AgCoO$_2$, (all at $T$=295K), and (d) LaCoO$_3$ at 20K.}
\end{figure}

In order to provide a reference spectrum with \emph{short Co--O bond lengths} but without non-local $1s3d$ transitions, we studied a low-spin Co(III) molecular compound with an octahedral CoO$_6$ core, where \emph{no neighboring Co} is present. The RXES of the selected Co(acac)$_{3}$, displayed in Fig.\ \ref{fig4}a, shows a single quadrupolar peak (QP) in the pre-edge, as a single $1s3d$ quadrupolar transition to a $^2E_g$ XAS final state is expected for a low-spin Co(III), which has a $t_{2g}^6e_g^0$ ground state.\cite{Note2Eg} (The vertical tail is a result of weak $2p^53d^7$ multiplet interactions.) Thus in the two cases presented so far, where Co--O--Co' mixing is weak or not possible, the pre-edge can be fully understood considering  quadrupole transitions with many-body multiplet effects.

\begin{figure}  
\includegraphics[width=8.2cm]{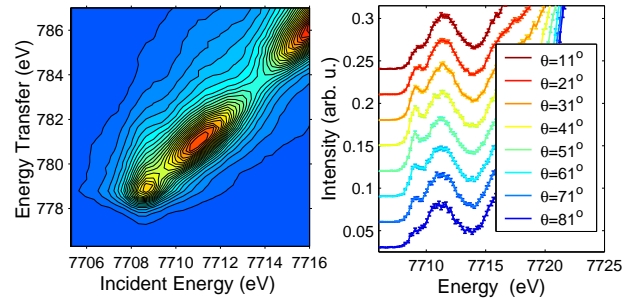}
\caption{\label{fig5} (Color online) Left panel: $1s2p$ RXES on EuCoO$_3$. Right panel: Angle-resolved XAS-like diagonal cross section of the RXES plane (taken at constant emission energy, as described in Ref.  \onlinecite{hamalainen1991}) on a single crystal. $\theta$ is the angle between the incident x-ray beam and the $c$ axis of the crystal.}
\end{figure}

In low-spin cobalt(III)-oxides Co($4p$)--O($2p$)--Co'($3d$) mixing is possible, and due to the short Co--O bond length ($\approx$ 1.92\AA) considerable non-local $1s3d$ dipole contributions can be expected. In Figs.\ \ref{fig4}b-- \ref{fig4}d and \ref{fig5} we present the RXES spectra of such oxides, including LiCoO$_2$, AgCoO$_2$, LaCoO$_3$ (at $T$=20~K), and EuCoO$_3$. The low-spin state and the similar first-shell structure of Co$^{3+}$ are confirmed by the very similar Co $L_{2,3}$ XAS spectra.\cite{haverkort2006,hu2004,lee2005}
Indeed, in all cases we can observe a second feature in the pre-edge region (\textit{hereafter:} DP), centered about 2.5 eV higher to QP. (This energy shift stems from the different core-hole screening, as discussed above.)

In order to obtain further support for the non-local $1s3d$ nature of DP, we can now study its behavior as a function of the linking of the CoO$_6$ octahedra, whose  The CoO$_6$ octahedra are connected at the corners in EuCoO$_3$ and in LaCoO$_3$, while in LiCoO$_2$ and in AgCoO$_2$ they are edge-sharing. Since the Co--O--Co angles are different in these compounds, different DP intensities are expected if they are indeed related to the oxygen-mediated $4p$-$3d$ mixing with the neighboring Co. In the edge-sharing geometry the Co--O--Co bond angle is close to 90$^\circ$ (LiCoO$_2$: 94.2$^\circ$,\cite{orman1984} AgCoO$_2$: 97.2$^\circ$ \cite{seshadri1998}). Consequently, the two neighboring Co$^{3+}$ ions cannot overlap with the same $2p$ orbital of the bridging oxygen. In the corner-sharing case, the bond angle is closer to linear (EuCoO$_3$: 152.9$^\circ$, LaCoO$_3$ ($T$=20K): 162.96$^\circ$,\cite{radaelli2002}) thus the same oxygen orbital can hybridize with both Co atoms. In fact, as we can see in Figs.\ \ref{fig4} and \ref{fig5}, this is the observed trend: in the spectra of the edge-sharing oxides the DP intensity is lower, while it dominates the spectra of the corner-sharing ones. In general, the DP intensity increases with the Co--O--Co bond angle.   

It is also worth noting that QP and DP show different dispersion with respect to the incident photon energy.4e While the first appears at a certain energy as a distinct resonance due to a transition to a narrow level, the second one has an elongated structure along the diagonal of the plane. Previous works have also reported elongated structures with different slopes; however, these usually appear because separate peaks merge due to the lack of resolution.\cite{degroot2005jpcb} In our case no fine structure of this broad feature was observed with 0.3 eV resolution. Moreover, this second feature lines up with the diagonal of the plane, suggesting that a portion of the excitation energy can be transferred to the kinetic energy of the promoted electron, which is expected for a transition to bands. Thus we can conclude that this second feature has all the characteristics to assign it to non-local $1s3d$ excitations. The different nature of these features is revealed by their angular dependence,\cite{brouder1990} as it is displayed in the right of Fig.\ \ref{fig5}: when changing the angle between the $c$ axis of the sample and the incident beam, the intensity of the first peak shows strong variation, suggesting an angular period of 90$^\circ$ (which indicates quadrupolar excitations to $3d$ orbitals), while the second feature changes little in the studied angular range. Consequently, the assignment of QP as the transition Q in Fig. 1 is confirmed, while DP has all characteristics of a non-local $1s3d$ dipolar peak D'.

Based on these observations we argue that the intensity of the second pre-edge peak, which we attribute to the non-local $1s3d$ dipolar peak, correlates with the factors that influence the $4p$-O-$3d$ intersite hybridization: the bond lengths and the bond angle. Its unusual dispersion suggests a band-like character, and its polarization dependence shows different multipolarity than that of the normal quadrupolar transition. The observed 2.5 eV shift from Q is understood as arising from the different screening effects. (Curiously, a shift of about 2.5 eV were also found in the pre-edges of TiO$_2$ and of La$_2$CuO$_4$ for the similar features.) 

It is noted that this non-local transition is different from the non-local screening in $2p$ X-ray Photoemission (XPS)\cite{vanVeenendaal1993prl,vanVeenendaal2006prb}. In $2p$ XPS the non-local screening involves an oxygen mediated $3d$--O--$3d$ hybridization, while in the non-local peak in the preedge it involves an oxygen mediated $4p$-O-$3d$ hybridization.
The main K edge involves dipole transitions from $3d^6$ to $1s^1 3d^6 4p^1$. The charge-transfer channels of the main edge are very similar to those of $1s$ XPS and $2p$ XPS, because the excited electron is essentially a delocalized state. As yet these channels are not distinguishable in the experiment and their analysis would need detailed correlated band structure calculations. In principle, the $3d3d$ non-local charge-transfer channel is also part of the charge-transfer channels of the main edge. This non-local screening channel is also observed in both soft and hard X-ray RIXS with a few eV energy transfer.\cite{collart2006,duda2006}

In systems with inversion symmetry, for example octahedral systems, the quadrupole and dipole states do not mix. The 2.5 eV energy difference can then be understood as the energy difference between the self-screened $1s^1 3d^7 4p^0 (3d^6)$ state and the bonding combinations of the three dipole final states. In principle there should be two additional states similar to non-local transitions and the poorly screened peak in 2p XPS.
This energy difference of 2.5 eV thus permits the experimental separation of the two overlapping different empty $3d$ states, as seen in Fig. \ref{fig1}c.

A relevant immediate consequence of our results concerns the spin state studies of cobaltates. LaCoO$_3$ exhibits a temperature-induced spin-state transition,\cite{haverkort2006,vanko2006prb} which in a localized model results in decreasing $t_{2g}$ and increasing $e_g$ population, $t_{2g}^{6-x}e_{g}^{x}$. With increasing temperature an increasing QP and a decreasing DP intensity were observed in the pre-edge region,\cite{vanko2006prb, medarde2006} which is in apparent agreement with the expected redistribution of the $3d$ electrons. Because of this, QP and DP were interpreted in the single-particle model as quadrupolar transitions to the $t_{2g}$ and $e_g$ levels, respectively. From the foregoing it is clear that this interpretation does not stand, and the spectral variations must arise from the temperature effects on the hybridization, which are indirectly related to the spin state (i.e., thermal expansion and diminishing mixing of metal ions of different spin states). This points out the importance of understanding the nature of the pre-edge features before such spectral features are interpreted; the current work shows how RXES can provide the necessary insights.

In summary, we report clear evidence for the theoretically expected low-lying empty states that stem from oxygen-mediated $4p$-$3d$ intersite mixing in cobalt oxides. High-resolution RXES applied to compounds built up by CoO$_6$ clusters reveals strong non-local $1s3d$ dipole excitations to these states in the Co $1s$ x-ray absorption preedge, which are identified by their energy offset, dispersion and angular dependence, and by their sensitivity to bond length and second-shell effects. Non-local $1s3d$ excitations are absent and the preedge can be fully described by $1s3d$ quadrupole transitions when the Co($4p$)--O--Co($3d$) hybridization is weak or no adjacent Co is present. Therefore, the approach presented provides an effective probe of the oxygen-mediated intersite $p$-$d$ mixing.
These results have important consequences on the understanding of the pre-edges (and thus the lowest-lying excitations) of highly correlated transition metal compounds with short metal-ligand distances. In these, typically high-valent systems, the usual interpretation of quadrupole pre-edge plus dipole edge fails and the dominant pre-edge structure can be attributed to the non-local $1s3d$ dipole feature.
This new interpretation of the pre-edges has far reaching implications within condensed matter physics, but also in many applied fields. In earth sciences and materials science, for example, when high-valent oxides and chalcogenides are investigated, transition metal $4p$-$3d$ intersite hybridization must be considered when electronic or magnetic properties or thermodynamic stability are being investigated, and, similarly, this effect must be considered in biology and chemistry for enzymes or bi- or polynuclear coordination compounds containing metal ions linked by a single atom of a bridging ligand.

We thank Pieter Glatzel, Giulio Monaco, and Keijo H{\"a}m{\"a}l{\"a}inen for discussions, Roberto Verbeni for making the analyzer crystal, and Christian Henriquet for invaluable technical assistance. 
This work was partly supported by the Hungarian Scientific Research Fund (OTKA) under contract No. K 72597. G.V. acknowledges support from the Bolyai J{\'a}nos Fellowship of the Hungarian Academy of Sciences.


\begin{thebibliography}{30}
\bibitem{Note2Eg} In fact, the quadrupolar transition leads to $1s^13d^{7}$ intermediate states, where a multiplet calculation shows that ~98\% of the intensity goes to the lowest energy $3d^{7}$ state of $^2E_g$ symmetry. The presence of effectively a single peak offers an optimal case to look for extra features in the pre-edge region.
\end{thebibliography}
\end{document}